\documentclass[iop,apj]{emulateapj} \usepackage{apjfonts} \newif\ifdraft \drafttrue \newif\ifpre \pretrue
\usepackage{xcolor}
\usepackage{url}
\usepackage{hyperref}
\hypersetup{
            colorlinks=true,
            linkcolor=black,
            filecolor=black,
            citecolor=black,
            urlcolor=blue
           }
\bibpunct{(}{)}{,}{a}{}{,}
\makeindex
\citeindextrue

\newcommand{\getlength}[1]{\ifx#1\end \let\next=\relax
            \else\advance\count255 by1 \let\next=\getlength\fi \next}
\newcount\switch
\newcommand{\Endmat}{\ifnum\switch=0$\fi}
\newcommand{\ifnularg}[1]{ \count255=0 \getlength#1\end \ifnum\count255=0 }
\newcommand{\ifm}{\makebox{}\ifmmode}

\long\def\ifundefined#1#2#3{\expandafter\ifx\csname
  #1\endcsname\relax#2\else#3\fi}
\newcommand{\beq}   { \begin{eqnarray} }
\newcommand{\eeq}[1]{ \ifnularg{#1} end{eanarray} \else
                      \label{#1}\end{eqnarray}    \fi }
\newcommand{\eeqn}{\nonumber\end{eqnarray}}
\newcommand{\Frac}[2]{\frac{\displaystyle\strut #1}{\displaystyle\strut #2} }
\newcommand{\ntab}[2]{ \multicolumn{1}{#1}{#2} }

\newcommand{\dss}{\displaystyle}
\newcommand{\vex}{\vspace{1ex}}
\newcommand{\pz}{\phantom{0}}
\newcommand{\eff}{\rm eff}
\newcommand{\SNR}{\rm SNR}
\newcommand{\Ac}{\mbox{$A_{\rm corr}$}}
\newcommand{\Apc}{\mbox{$A^p_{\rm corr}$}}
\newcommand{\Fc}{\mbox{$F_{\rm corr}$}}
\newcommand{\Fsc}{\mbox{$F^s_{\rm corr}$}}
\newcommand{\Frc}{\mbox{$F^r_{\rm corr}$}}
\newcommand{\lp}{ \left(  }
\newcommand{\rp}{ \right) }
\newcommand{\PIMA}{$\cal P\hspace{-0.067em}I\hspace{-0.067em}M\hspace{-0.067em}A$ }

\newcommand{\Number}[1]{\ifnum#1<10\relax0\number#1\else\number#1\fi}
\newcommand{\isodate}{
\count151=\time
\divide\count151 by 60
\count151=\count151
\multiply\count151 by 60
\count152=\time
\advance\count152 by -\count151
\divide\count151 by 60
\count152=\count151
\multiply\count151 by 60
\count153=\time
\advance\count153 by -\count151
\Number{\year}.\Number{\month}.\Number{\day}--\Number{\count152}:\Number{\count153}
}
\definecolor{Dred}{rgb}{0.312,0.070,0.070}
\definecolor{Dblue}{rgb}{0.070,0.070,0.312}
\definecolor{Dgreen}{rgb}{0.070,0.312,0.070}
\definecolor{Db}{rgb}    {0.050,0.0,0.320}

\newcommand{\Blb}[1]{\textcolor{Dblue}{\bf #1}}

\newcounter{note}
\let\oldmarginpar\marginpar
\renewcommand\marginpar[1]{\-\oldmarginpar[\raggedleft\footnotesize #1]%
{\raggedright\footnotesize #1}}

\ifdraft
    \newcommand{\web}[1]{\Blb{\url{#1}}}
  \else
    \newcommand{\web}[1]{\url{#1}}
\fi

\shorttitle{The KCAL VERA 22~GHz calibrator survey}
\shortauthors{Petrov et al.}

\begin{document}

\title{The KCAL VERA 22~GHz calibrator survey}

\author{L. Petrov\altaffilmark{1}}
\affil{Astrogeo Center, Falls Church, USA}
\email{Leonid.Petrov@lpetrov.net}

\author{M. Honma\altaffilmark{2} and S.~M. Shibata\altaffilmark{2}}
\affil{National Astronomical Observatory of Japan, Mitaka, Japan}

  \submitted{}
  \received{June 13, 2011}
  \revised{July 13, 2011}
  \accepted{July 17, 2011}

\ifpre
  \submitted{}
  \received{2011, Ocober 06}
  \revised{2011, November 14}
  \accepted{2011, Novemeber 15}
\else
  \paperid{AJ-10787R}
\fi

\begin{abstract}


   We observed at 22~GHz with the VLBI array VERA a sample of 1536 sources
with  correlated flux densities brighter than 200~mJy at 8~GHz. One half
of target sources has been detected. The detection limit was around 200~mJy.
We derived the correlated flux densities of 877 detected sources in three
ranges of projected baseline lengths. The objective of these observations
was to determine the suitability of given sources as phase calibrators for
dual-beam and phase-referencing observations at high frequencies.
Preliminary results indicate that the number of compact extragalactic
sources at 22~GHz brighter than a given correlated flux density level
is twice less than at 8~GHz.


\end{abstract}
\keywords{Quasars and Active Galactic Nuclei}

\section{Introduction}

  Currently, very long baseline interferometry (VLBI) astrometry 
is the best tool to measure distances and motions of sources located 
at kpc scale and hence, to explore the structure of the Milky Way in 
the Galactic scale. For instance, Japanese VERA project (VLBI Exploration 
of Radio Astrometry; \citet{r:vera00}) has been conducting astrometric 
monitoring of positions of Galactic maser sources with respect to reference 
compact extragalactic objects, yielding handful measurements of 
parallaxes and proper motions of maser sources (e.g., see recent 
PASJ special issue for VERA, such as \citet{r:hon11,r:nag11} and 
others). The Very Long Baseline Array (VLBA) is actively used for 
astrometry of Galactic maser sources (\citep[e.g.,][]{r:reid09} 
and the currently ongoing Bar and Spiral Structure Legacy (BeSSeL) survey)
and the European VLBI Network (EVN) conducts astrometric observations 
of methanol maser sources \citep[e.g.,][]{r:ryg10a}.

  In order to measure parallax and proper motion of a radio source
at kpc scales, it is observed in the phase-referencing mode by frequent 
switching pointing between the target and a calibrator source. This technique
significantly reduces phase variations caused by tropospheric fluctuations.
To do this effectively, calibrators must be located close to target sources, 
typically within 1--2\degr\ separation. This requires a high density
of calibrator sources in the sky, and hence, there is still
a strong demand for finding many calibrator sources.

  To date, there have been several massive surveys of compact
calibrators such as VCS (VLBA Calibrator Surveys), \citet{r:vcs6}
and references therein), the LCS (Long Baseline Array Calibrator Survey) 
for the southern hemisphere \citep{r:lcs1}, VIPS (VLBA Image and Polarization
Survey) \citep{r:vips,r:astro_vips}, and several 
ongoing programs: the program of study the {\it Fermi} active galaxy nuclea
(AGNs) at parsec scales{\footnote{\web{http://astrogeo.org/faps}}}
(Kovalev et al. (2011), paper in preparation), the program of observing
radio-loud 2MASS (Two Micron All Sky Survey) 
galaxies{\footnote{\web{http://astrogeo.org/v2m}}} \citep{r:con11}, 
the program of observing optically bright 
quasars \citep{r:bou08,r:bou11,r:obrs1}, and the recent VLBA calibrator 
search for the BeSSeL survey \citep{r:br145a}.

  Together with regular geodetic VLBA observations of 1000 sources
\citet[the RDV program][]{r:rdv}), by June 2011 positions of 6455 sources
at a milliarcsecond level of accuracy were derived from analysis of these
massive surveys. The sources turned out compact enough to be detected
with VLBI, i.e. they have a core of mas scale. However, these surveys were 
in most cases conducted in relatively low frequencies such as 2 (S-band), 
5 (C-band) or 8~GHz (X-band), at which the telescope performance is the best.
On the other hand, recent VLBI maser astrometry is often done at 
frequencies higher than 10~GHz. For instance, VERA's main bands 
are 22 (K-band) and 43~GHz (Q-band) for H$_2$O and SiO maser sources. 
Maser astrometry with VLBA is mainly conducted at 12~GHz for CH$_3$OH masers 
and 22~GHz for H$_2$O masers. Therefore, calibrator
information at high frequencies (such as K and higher bands)
is of great importance for on-going and future astrometric
observations. Compact calibrators which are cores of radio bright
AGNs have a wide variety of their spectra: for the majority of
sources the correlated flux density decreases with the frequency,
although some sources may have spectra growing with frequency
or peaking within the GHz regime. Hence, the extrapolation of the 
correlated flux densities from S and X band to 22 or 43~GHz is highly
unreliable. For successful phase-reference or dual-beam observations,
the correlated flux density should be known with accuracy at least 30\%
in order to correctly predict the signal-to-noise ratio (\SNR).
Therefore, it is necessary to conduct a systematic
survey of K-band flux densities for the compact sources which
were already detected in S and X bands.

  We have identified $\sim\!\!2000$ sources previously observed with VLBI
with $\delta > -30\degr$ with correlated flux densities $>200$~mJy
at X-band at baselines longer than 900~km. Analysis of 
the dependence of the number of sources $N$ with the 
correlated flux density exceeding $S$ as a function of $S$ suggests that 
this sample is complete at the 95\% level (Kovalev 2010, private 
communication). Of these sources, 511 have been previously observed 
in large K-band surveys: VERA Fringe Search Survey \citep{r:vera_fss}, 
KQ survey \citep{r:kq}, VLBI Galactic plane survey (VGaPS) \citep{r:vgaps}, 
in the EVN Galactic plane survey (EGaPS) \citep{r:egaps}, and their 
correlated flux densities at 22~GHz have been measured. 
The K-band brightness of other objects was not known.

  We conducted a dedicated survey of remaining 1536 sources
at 22~GHz with VERA in the K-band Calibrator Survey (KCAL) campaign.
The goal of these observations was to check their detectability at K-band
and to measure the correlated flux densities of detected sources
at baselines 1000--2000~km.

  The first objective of this campaign was to provide a complete list
of calibrators suitable for VERA observations of faint targets. According
to our prior observations, the detection limit of the VERA network for
2 minutes of integration time is around 200~mJy, depending on weather
conditions. Therefore, the list of sources observed in this and the
previous K-band surveys is expected to approach the completeness at 
the 200~mJy level, provided the spectra of compact cores are flat or falling. 
According to \citet{r:mass10} who analyzed simultaneous ATCA spectra 
at 4.8, 8.4 and 20~GHz, the share of sources with growing spectra that 
may be missed in our sample does not exceeded 8\%.

  The second objective of this campaign is to collect information
for a population analysis of a large complete sample. In particular,
the analysis of the dataset that combines existing and new data will
help to answer the question what is the distribution of spectral indexes
of the core regions and the source compactness at high frequencies,
whether the spectral index at parsec scales is systematically different
than the spectral index at kiloparsec scales, and whether the compactness
at K-band is systematically different than the compactness at X and
S bands.

  In this paper we present results of the survey. In section
\ref{s:obs} we describe the observations, their design and scheduling.
In section \ref{s:anal} we discuss analysis technique. The catalogue
of correlated flux densities of detected sources accompanied with analysis
of flux density uncertainties is presented in section \ref{s:cat} followed
by concluding remarks that are given in section \ref{s:sum}.

\section{Observations}
\label{s:obs}

  Observations were carried out using the network of four 20 meter 
antennas VERA at K-band. The primary task of the array is to perform 
parallax measurement of maser sources. In order to maximize the throughput 
of the instrument, observing time for the KCAL experiments was allotted 
in blocks that fill gaps between parallax measurement observing sessions 
or during periods of time when one of the antennas was under maintenance.


  A monthly observing plan for VERA parallax measurements was usually 
finalized by at least one week before the beginning of the month. 
When there were suitable gaps for KCAL experiments and there were enough  
magnetic tapes in the Mitaka correlation center, we ran calibrator survey
experiments during these gaps. The parallax measurement requires 
participation of each of four stations of VERA in order to achieve required 
astrometric accuracy. If any station, other than Ogasawara, could 
not join regular observations because of maintenance or instrumental 
problems, the KCAL experiments were also scheduled during that time with 
three stations.
%

  The left circular polarization in the 21.97--22.47 GHz band was received,
sampled with 2~bit quantization, and filtered using the VERA digital
filter \citep{iguchi2005} before being recorded onto magnetic tapes. The
digital filter split the data within the 500~MHz band into 16 frequency
channels of 16~MHz width each, equally spaced with 16~MHz wide gaps.

\subsection{Scheduling}

  Scheduling software {\sf sur\_sked} selected sources from the pool
of candidate objects in a sequence that minimizes slewing time.
At a given experiment, each source was observed in one scan of
120~seconds long. Every 30~minutes a scan of a strong source with
the brightness distribution map produced from VLBA observations
under the KQ observing campaign \citep{r:kq} was inserted in
the schedule. The purpose of including these scans in the schedule
was to compare our measurements of the correlated flux densities
of sources with known images considered as the ground truth in order
to evaluate gain corrections. The target sources which were observed
in one scan were returned to the pool for scheduling in the second
scan in following experiments.

  In total, 36 experiments were scheduled. However, six experiments were
canceled for various reasons, in three observing sessions two stations
either failed or did not observe; these experiments were excluded
from analysis. The dates and durations of the 27 VLBI experiments under
the KCAL program over the period 2007--2009 that were used in the analysis
are shown in Table~\ref{t:obs}.

\begin{table}[h]
  \caption{\ifpre \rm \fi
           Dates and durations of experiments. Only those
           experiments that were used in the final analysis
           are shown. Station abbreviations: Ir for Iriki,
           Is for Ishigakijima, Mz for Mizusawa, Og for Ogasawara.}
  \label{t:obs}
  \par\medskip\par
  \begin{tabular}{l l r l}
    \hline
    Exp ID    & Date       & Dur (h) & Network  \\
    \hline
    kcal\_01  & 2007.05.28 &  5.3 & Ir Is Mz Og \\
    kcal\_02  & 2007.05.30 &  5.8 & Ir Is Mz Og \\
    kcal\_03  & 2007.05.31 &  3.9 & Ir Mz Og    \\
    kcal\_04  & 2007.08.24 &  3.8 & Ir Is Mz Og \\
    kcal\_05  & 2007.08.24 & 14.3 & Ir Is Mz Og \\
    kcal\_06  & 2007.08.25 &  6.8 & Ir Is Mz Og \\
    kcal\_07  & 2007.11.18 &  4.8 & Ir Is Mz Og \\
    kcal\_09  & 2007.11.23 &  4.5 & Ir Is Mz Og \\
    kcal\_10  & 2007.12.10 &  5.8 & Ir Is Mz Og \\
    kcal\_11  & 2007.12.12 &  2.6 & Ir Mz Og    \\
    kcal\_12  & 2007.12.12 &  2.8 & Ir Mz Og    \\
    kcal\_15  & 2007.12.19 &  5.8 & Ir Is Mz Og \\
    kcal\_16  & 2007.12.20 &  3.8 & Ir Is Mz Og \\
    kcal\_17  & 2007.12.21 &  3.8 & Ir Is Og    \\
    kcal\_18  & 2007.12.22 &  2.5 & Ir Is Og    \\
    kcal\_19  & 2007.12.22 &  3.8 & Ir Is Mz Og \\
    kcal\_23  & 2008.02.29 &  6.8 & Ir Is Mz Og \\
    kcal\_24  & 2008.06.03 &  2.1 & Ir Is Mz Og \\
    kcal\_25  & 2008.06.11 &  5.2 & Ir Is Mz Og \\
    kcal\_27  & 2008.10.06 & 15.8 & Ir Is Mz Og \\
    kcal\_29  & 2008.10.12 &  3.6 & Ir Is Mz Og \\
    kcal\_30  & 2008.11.11 &  2.3 & Ir Is Mz Og \\
    kcal\_31  & 2008.11.16 &  2.1 & Ir Is Mz Og \\
    kcal\_32  & 2008.11.14 &  4.0 & Ir Is Mz Og \\
    kcal\_33a & 2009.03.13 &  7.9 & Ir Is Mz    \\
    kcal\_33c & 2009.03.21 &  3.9 & Ir Is Mz   \\
    kcal\_33d & 2009.03.22 &  8.8 & Ir Is Mz    \\
    \hline
  \end{tabular}
\end{table}

  The scheduling goal of the campaign was to have each target 
source observed in two experiments, one scan in each observing session. 
Due to the nature of scheduling in a fill-in mode, it turned out difficult 
to reach this goal. As it seen from Table~\ref{t:staobs}, 1/3 of the 
sources were observed in one scan. In total, 1536 target sources 
were observed for 143 hours. The antennas spent 71\% time on target
sources. Remaining time was spent for observing the
amplitude calibrators and for slewing.

\begin{table}[h]
  \caption{\ifpre \rm \fi
           Statistics of the number of scans per observed source.
           The first columns shows the number of scans, the second
           table shows the number of target sources which had that
           number of scans. The last column shows the share
           of sources from the target list which had that number of scans.
          }
  \label{t:staobs}
  \par\medskip\par
  \begin{center}
    \begin{tabular}{ c @{\quad} r @{\quad} r}
      \hline
      \# scans & \# obs & Share   \\
      \hline
             1 &   530  &  35\%   \\
             2 &   633  &  41\%   \\
             3 &   267  &  17\%   \\
             4 &   102  &  5\%    \\
             5 &     4  &  0.2\%  \\
      \hline
    \end{tabular}
  \end{center}
\end{table}

\section{Data analysis}
\label{s:anal}

\subsection{Fringe fitting}

   The data were correlated on the Mitaka FX correlator \citep{chikada1991}.
Correlation output was written in the FITS-IDI format. Consecutive analysis
was performed with computer program 
\PIMA$\!$\footnote{Available at \web{http://astrogeo.org/pima}}. The procedure 
of data analysis is described in detail in \citet{r:vgaps}. Here only 
a brief outline is given. After applying correction of fringe amplitude 
for digitization, the spectrum of the cross-correlation function was presented 
as a two-dimensional array with the first dimension running over frequency
channels and the second dimension running over time. The two-dimensional
Fourier-transform of the spectrum over frequency and time cast
the spectrum of the cross-correlation function into the domain of group
delay and phase delay rate. A set of estimates of delays, phase delay
rates and fringe amplitude for a given scan at a given baseline is 
thereafter called observation. In the presence of the signal in the data,
the Fourier-transform of the cross-spectrum exhibits a sequence of peaks.
The amplitude of the major peak is proportional to the fringe amplitude
of the signal. The fringe fitting process locates the peaks and determines
the group delay, delay rate and fringe amplitude that correspond to
the main maximum of the Fourier-transform of the cross-correlation spectrum.

   In order to determine the detection threshold, first we have
to measure the noise level. To do this, we computed the ratios of fringe 
amplitudes to mean amplitudes of the Fourier-transform of the cross-correlation 
spectrum. That mean amplitude was computed by averaging 32768 randomly 
selected samples of the cross-spectrum Fourier-transform after iterative excluding 
the amplitudes that are greater than 3.5 times of the variance of amplitudes 
in the sample, in order to be sure that no samples with the signal were selected
by accident. This procedure ensures that the mean amplitude of the 
noise is determined with an accuracy no worse than 1\%.

   Even in the absence of the signal, the fringe fitting procedure will
find a peak, but the amplitude of this peak will not be related to the
fringe amplitude. The distribution of the achieved {\SNR}s consists of 
the contribution of the population of observations with signal detected and 
the population of observations without signal. The \SNR\ probability density 
in the absence of signal is described (for instance \citep{r:vgaps}):
\beq
    p(s) = \Frac{2}{\pi} \Frac{n_{\eff}}{\sigma_{\eff}} \, s \,
           e^{-\frac{s^2}{\pi}}
           \left( 1 - e^{-\frac{ s^2}{\pi}} \right)^{n_{\eff}-1} ,
\eeq{e:e1}
  where $n_{\eff}$ is the effective number of independent samples
and $\sigma_{\eff}$ is the effective noise variance.

\begin{figure}[h]
  \caption{\ifpre \rm \fi
           The left tail of the empirical distribution of the achieved
           \SNR\ from results of fringe fitting VERA data (filled circles) 
           and the fitted curve (thin line) of the theoretical 
           distribution for the case of no signal.
  }
  \label{f:prob}
  \par\medskip\par
  \ifdraft
     \includegraphics[width=0.46\textwidth]{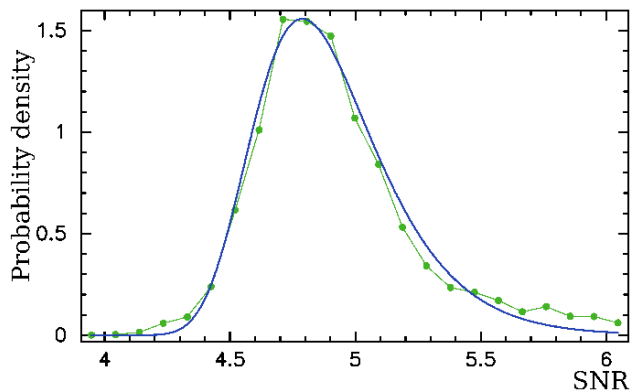}
  \else
     \includegraphics[width=0.46\textwidth]{kcal_prob.eps}
  \fi
\end{figure}

   In order to determine $n_{\eff}$ and $\sigma_{\eff}$, we 
computed the histogram of the achieved \SNR\ in the KCAL experiments 
in the range of [3.8, 6.5] (see Figure~\ref{f:prob}) and fitted it with 
the theoretical curve  $p(s)$ of the fringe amplitude distribution 
in the absence of the signal. The left tail of the \SNR\ histogram
is dominated by non-detected sources. The right tail is dominated by
detected sources. The breakdown occurs with SNR in a range of [5, 6.5].
There is some fraction of detected sources with the \SNR\ within the range 
of [5, 6.5], and they potentially may cause a bias in our estimates 
of $n_{\eff}$ and $\sigma_{\eff}$. We varied the range of {\SNR}s used for 
fitting and found that the estimates are stable at a level of $10^{-3}$, 
i.e. the bias is negligible.

  After determining $n_{\eff}$ and $\sigma_{\eff}$, we can find the
probability that an observation with a given \SNR\ belongs to the population
of observations without a signal, i.e. the probability of false detection, 
by integrating expression \ref{e:e1} over $s$, which can be easily done 
analytically. Specifically, we found that the probability of false detection 
is less than 0.001 when the $\SNR\ > 6.03$. We considered a source as detected
if the \SNR\ in at least two observations at different baselines of the same
scan was above the detection limit 6.03. In the absence of the signal, 
the probability of finding two peaks exceeding the threshold limit in data of 
different observations is in the range of $10^{-3}$ to $10^{-6}$ depending 
whether the errors are completely correlated or completely uncorrelated. 
In practical terms, this means that our catalogue may have no more than one 
or two falsely detected objects.

\subsection{Amplitude calibration}

  System temperatures including atmospheric attenuation were measured with
the chopper-wheel method \citep{r:ulich1976}. At the beginning of each scan,
a microwave absorber at ambient temperature was inserted just in front of
the feed horn, and the received total power was measured with a power meter. 
Using the measured total power for the blank sky and the absorber, 
the temperature scale automatically corrected for the atmospheric attenuation
was determined. We estimate the uncertainty in the temperature scale 
around 10\%, mainly due to the assumption that the ambient temperature 
is the same as the air temperature.

  The initial amplitude calibration was made by scaling fringe amplitudes,
determined with the fringe fitting process, by the measured system temperature
and dividing them by the antenna gain. Then, the  antenna gains 
were adjusted by comparing the calibrated fringe amplitudes 
of the observed calibrator sources with the correlated flux densities
predicted on the basis of their K-band brightness 
distributions\footnote{Available at \web{http://astrogeo.org/vlbi_images}
produced from analysis of observations from the KQ \citep{r:kq} and
VLBA Galactic Plane Surveys (VGaPS) \citep{r:vgaps}}:
\beq
   \begin{array}{lcl}
      F_{corr} & = & \left|
                   \dss\sum_j c_j(x,y) \,
                              e^{\frac{2\pi i\, f}{c}\, (u\, x + v\, y)}
                   \right| , \\
   \end{array}
\eeq{e:e2}
   where $c_i$ is the correlated flux density of the $j$th CLEAN component
with coordinates $x$ and $y$ with respect to the center of the image;
$u$ and $v$ are the projections of the baseline vectors onto the tangential
plane of the source.

  Then we built a system of equations for all observations of calibrators:
\beq
     \Fc = \sqrt{g_i \, g_j} \Ac
\eeq{e:e3}
   that relates the calibrated amplitude $\Ac$, gain corrections $g$
for stations $i$ and $j$ of a baseline, and the predicted correlated flux
of the amplitude calibrator. After taking logarithms from left and right
hand sides, we solved for average gains corrections for all stations using
the least squares (LSQ) technique. Then an iterative procedure of outliers 
elimination was performed. At each step of iterations, we computed 
the rms of the ratio of observed and predicted correlated flux densities.
We searched for the observation with the maximum by module logarithm of 
this ratio. If the ratio for that observation exceeded $3.5\sigma$ rms,
we excluded the observation from the system of equations and ran a new
LSQ solution. The process was repeated till no observations with
the maximum by module logarithm exceeding $3.5\sigma$ rms was found.

  The number of calibrators in each individual experiment varied.
On average, 9 calibrators were used for gain correction adjustment
in each experiment. If the model brightness distributions were perfect,
and gain corrections were stable over an experiment, calibration errors
would have been below the noise level. Several factors can degrade 
the quality of calibration using this method. First, the images 
of calibrator sources
were produced using observations at different sampling of spatial frequencies
than the analyzed observations. Computation of the predicted correlated
flux densities is equivalent to an interpolation of visibilities measured
in KQ and VGaPS VLBA campaigns to $u$ and $v$ baseline projections in the
KCAL experiments. Errors of this interpolation may be significant, except
for sources with very simple structure. Second, both source structure
and the peak brightness evolve with time. Since the time difference
in epochs between KQ, VGaPS and KCAL experiments is 2--6 years, the changes
in source brightness distribution may be significant. The sampling bias
and the source variability are expected to cause only random errors in gain,
but not a systematic bias. Some calibrator sources may become brighter,
some dimmer, but the average flux density of the ensemble should be rather
stable. Third, we assumed that gain corrections are constant over time
of an individual experiment since we do not have enough information for
modeling their time variability.

\section{The correlator flux density catalogue}
\label{s:cat}

%
%
%
%


   Since the data are too sparse to produce meaningful images, we
computed average correlated flux densities for detected sources in
three ranges of projected baseline lengths: 0--70 megawavelengths,
70--100 megawavelengths, and 100--250  megawavelengths, which corresponds
to lengths 0--955~km, 955--1365~km, and longer that 1365~km respectively.
The corresponding resolutions are $>3$~mas for the first range,
2--3~mas for the second range, and $<2$~mas for the third range.
The amplitudes were calibrated for gain corrections using the method
described in the previous section. This simplified method of correlated
flux density evaluation is an alternative to a rigorous imaging procedure
in the case when there are too few measurements.

   The catalogue of correlated flux densities of 877 observed sources, 
including 750 targets and 127 calibrators, is presented
in table~\ref{t:cat}. Objects with at least two detections are put in
the catalogue. Columns~1 and~2 show IAU and IVS source names. Column~3
shows the source status: C stands for an amplitude calibrator, blank stands
for a target object. Column~4 shows the number of experiments in which
a source was detected and column~5 shows the total number of detections.
Columns~6, 7, and~8 present the estimates of the average correlated flux
density in three ranges of the projected baseline lengths. Columns~9, 10,
and~11 show the estimates of correlated flux density uncertainty:
$ \sigma(\Fc) = \Ac \cdot \sqrt{0.2^2 + \frac{2}{\pi}\frac{1}{\SNR^2} }$.
Columns~12 and~13 show right ascensions and declinations. 
Value $-1.000$ in columns 6--11 indicates a lack of results for these 
baseline projections.

\ifpre
  \begin{deluxetable*}{l l @{$\;\;$} c @{$\;$} r @{$\;\;$} r @{\qquad} 
                      r r r @{\quad\qquad} r r r @{\quad\qquad} r r}
\else
  \begin{deluxetable} {l l @{$\;\;$} c @{$\;$} r @{$\;\;$} r @{\qquad} 
                      r r r @{\quad\qquad} r r r @{\quad\qquad} r r}
  \rotate
\fi
  \tablecaption{\ifpre \rm \fi
                The first 12 rows of the catalogue of correlated flux densities 
                of 877~sources that have at least two detections in VERA KCAL 
                observing campaign. 
                The table columns are explained in the text. 
                The full table is available in the electronic attachment. 
                \label{t:cat}
               }
  \tablehead{
             \multicolumn{13}{c}{ } \vspace{3ex} \\
             \multicolumn{2}{c}{\normalsize Source names}  &
             &
             \multicolumn{2}{c}{\normalsize Statistics}    &
             \multicolumn{3}{c}{\normalsize Corr. flux density}    &
             \multicolumn{3}{c}{\normalsize Errors of $F_{\rm corr}$}  &
             \multicolumn{2}{c}{\normalsize Source coordinates}     
             \\ [0.5ex]
             \ntab{c}{(1)}    &
             \ntab{c}{(2)}    &
             \ntab{c}{(3)}    &
             \ntab{c}{(4)}    &
             \ntab{c}{(5)}    &
             \ntab{c}{(6)}    &
             \ntab{c}{(7)}    &
             \ntab{c}{(8) \phantom{aaa}  } &
             \ntab{c}{(9)}    &
             \ntab{c}{(10)}   &
             \ntab{c}{(11) \phantom{aaa} } &
             \ntab{c}{(12)}   &
             \ntab{c}{(13)}
             \vspace{1ex}
             \\ 
             IAU name &
             IVS name &
             flag &
             \#Exp &
             \#Det &
             $F_{<70}$    &
             $F_{70-100}$ &
             $F_{>100}$   &
             $E_{<70}$    &
             $E_{70-100}$ &
             $E_{>100}$   &
             Right ascen  &
             Declination
            \\ 
                &
                &
                &
                &
                &
   \ntab{c}{Jy} &
   \ntab{c}{Jy} &
   \ntab{c}{Jy \phantom{aaa}} &
   \ntab{c}{Jy} &
   \ntab{c}{Jy} &
   \ntab{c}{Jy \phantom{aaa}} &
   \multicolumn{1}{l}{\pz h \pz m \pz s} &
   \multicolumn{1}{l}{\pz\pz $\degr \pz\pz {}' \pz\pz {}''$}
    }
    \startdata
    \vex \vex \vex \\
    J0001$+$1914 & 2358$+$189 &   &  1 &  4 &  0.221 &  0.322 &  0.216 &   0.055 &  0.070 &  0.051 &  00 01 08.62   & $+$19 14 33.8 \\ [1pt]
    J0005$+$3820 & 0003$+$380 &   &  2 &  8 & -1.000 &  0.608 &  0.526 &  -1.000 &  0.136 &  0.116 &  00 05 57.17   & $+$38 20 15.1 \\ [1pt]
    J0006$-$0623 & 0003$-$066 &   &  2 &  9 &  1.027 &  1.120 &  1.212 &   0.104 &  0.205 &  0.176 &  00 06 13.89   & $-$06 23 35.3 \\ [1pt]
    J0008$+$6837 & 0005$+$683 &   &  1 &  2 & -1.000 &  0.353 & -1.000 &  -1.000 &  0.080 & -1.000 &  00 08 33.47   & $+$68 37 22.0 \\ [1pt]
    J0010$+$1058 & IIIZ$W$2   & C &  2 &  6 & -1.000 &  1.193 &  1.442 &  -1.000 &  0.239 &  0.289 &  00 10 31.00   & $+$10 58 29.5 \\ [1pt]
    J0010$+$1724 & 0007$+$171 &   &  1 &  2 & -1.000 &  0.340 &  0.266 &  -1.000 &  0.078 &  0.060 &  00 10 33.99   & $+$17 24 18.7 \\ [1pt]
    J0010$-$2157 & 0008$-$222 &   &  1 &  2 & -1.000 &  0.236 & -1.000 &  -1.000 &  0.052 & -1.000 &  00 10 53.64   & $-$21 57 04.2 \\ [1pt]
    J0011$+$7045 & 0008$+$704 &   &  2 &  8 &  0.440 &  0.405 &  0.542 &   0.043 &  0.088 &  0.064 &  00 11 31.90   & $+$70 45 31.6 \\ [1pt]
    J0012$-$3954 & 0010$-$401 &   &  1 &  3 &  0.505 &  0.494 & -1.000 &   0.120 &  0.103 & -1.000 &  00 12 59.90   & $-$39 54 26.0 \\ [1pt]
    J0013$+$4051 & 0010$+$405 &   &  2 &  7 &  0.531 &  0.534 &  0.492 &   0.065 &  0.119 &  0.115 &  00 13 31.13   & $+$40 51 37.1 \\ [1pt]
    J0013$-$0423 & 0011$-$046 &   &  2 &  7 &  0.552 &  0.406 &  0.445 &   0.071 &  0.094 &  0.101 &  00 13 54.13   & $-$04 23 52.2 \\ [1pt]
    J0017$+$5312 & 0015$+$529 &   &  1 &  2 & -1.000 &  0.270 & -1.000 &  -1.000 &  0.067 & -1.000 &  00 17 51.75   & $+$53 12 19.1 \\ [1pt]
    \enddata
\tablecomments{ \rm
                Table~\ref{t:cat} is presented in its entirety in the electronic
               edition of the Astronomical Journal. A portion is shown here
               for guidance regarding its form and contents. 
              }
\ifpre
  \end{deluxetable*}
\else
  \end{deluxetable}
\fi

  Of 1536 observed sources, including both targets and calibrators,
407 were not detected at all and 252 were detected only in one observation.
The detections from the latter group were considered unreliable and were not
included in the catalogue.

\subsection{Error analysis}

   Errors in correlated flux density estimates are due to 1)~the thermal
noise in estimates of fringe amplitude; 2)~the uncertainties in system
temperature measurements; 3)~the uncertainties in antenna gain measurement;
4)~the sampling bias in predicted correlated flux densities of calibrators;
5)~the variability of calibrator sources.

  The uncertainty due to the thermal noise can be easily evaluated as
$\sqrt{2/\pi <\!\!a_n\!\!>/a}$, where $<\!\!a_n\!\!>$ is the average
amplitude of the noise computed by the fringe fitting procedure, 
and $a$ is the fringe amplitude.
As we already mentioned, the uncertainty in system temperature measurement 
is around 10\%. The aperture efficiency of VERA antenna is measured every 
year and known within 10\% accuracy (see VERA status 
report\footnote{Available at \web{http://veraserver.mtk.nao.ac.jp/}}). 
We assume these two uncertainties uncorrelated, and therefore, these two 
factors would introduce an uncertainty of the a~priori gain calibration 
at $\sim\!14\%$ level.

  Since on average, nine amplitude calibrators were used for gain
adjustments, this redundancy can be exploited for evaluation the
gain correction uncertainties. We computed the average and the root mean
square (rms) of the residual mismatches between observed correlated flux
densities of calibrators after applying gain corrections $\Apc$ from the
LSQ fit and the predicted correlated flux densities from the brightness
distributions:
\beq
    \begin{array}{lcl}
       \mbox{\rm Avr} & = &
             \lp \dss\prod\limits_{i} \Frac{\Fc_{,i}  \sqrt{g_1 g_2}}{\Apc_{,i}}
             \rp^{1/n}
                      \vspace{1ex} \\
       \mbox{\rm Rms} & = &
             \sqrt{ \Frac{ \dss\sum_i \lp  \Frac{\Fc_{,i} \sqrt{g_1 g_2}}
                                           {\Apc_{,i}} - 1
                                      \rp^{2}
                         }
                         {n}
                  } .
    \end{array}
\eeq{e:e4}

  We found Avr = 0.994 and Rms = 0.21. The first statistics describes the
systematic bias and the second statistics is the measure of the contribution
of uncertainties in gain adjustments on the uncertainty of our estimate
of the correlated flux density.

%
%

  In order to evaluate the representativeness of this statistics,
we computed the median correlated flux densities in three ranges of
projected baseline lengths of two experiments of the 24~GHz VLBA
VGaPS campaign using two methods: 1)~rigorous self-calibration imaging
and 2)~the same simplified method used for processing KCAL experiments.
In order to closely mimic analysis of the KCAL experiments, we used for our
tests the brightness distributions from the KQ campaign made at epochs
at least one year prior to observations. We got Avr = 0.996 and
Rms = 0.24. Then we computed the rms of the scatter of the ratios of
the correlated flux density $\Fsc$ determined by the simplified method to
the flux density $\Frc$ determined by the rigorous method:
rms = $\sqrt{ \sum_i (\Fsc_{,i}/\Frc_{,i} - 1)^2}$. We found the rms
equal to 0.15. Considering the brightness distributions from the
self-calibration analysis procedure as the ground truth, we
conclude that the accuracy of the median correlated flux density obtained
by the simplified method is at a level of 15\% for the VGaPS campaign.
Thus, the Rms statistics give us rather an upper limit of gain errors.

  Another way to evaluate the average uncertainty of correlated flux 
densities is to compute the rms of the scatter of ratios of the correlated 
flux densities of a given source with respect to the mean value for all 
the KCAL sources which have three or more observations. We got the value 
of the rms 0.20, which is close to the Rms statistics. Therefore, 
we conclude that the average uncertainty of calibration error is 20\%. 
Since the uncertainty in fringe amplitude caused by the thermal noise and 
calibration errors are independent, we compute the multiplicative 
uncertainty of reported correlated flux density as a sum of these 
two contributions in quadrature: 0.2 and 
$\sqrt{\frac{2}{\pi}}\frac{1}{\SNR}$.

\section{Concluding remarks}
\label{s:sum}

   We observed with VERA at 22~GHz a subset of the complete sample
of continuum compact extragalactic sources with correlated flux densities
$> 200$~mJy at X-band at declinations $>-30\degr$. The subset excluded
the sources previously detected at K-band at large VLBA and VERA surveys.
Of 1536 target sources, approximately one half has been detected.
The errors of the correlated flux densities are a level of 20\%.

\begin{figure}[h]
  \caption{\ifpre \rm \fi
           The distribution of the correlated flux densities at baseline
           projection lengths longer than 100 megawavelengths. The last
           bin of the histogram has all the sources with correlated
           flux density $>2$ Jy.
  }
  \label{f:distr}
  \par\medskip\par
  \ifdraft
     \includegraphics[width=0.46\textwidth]{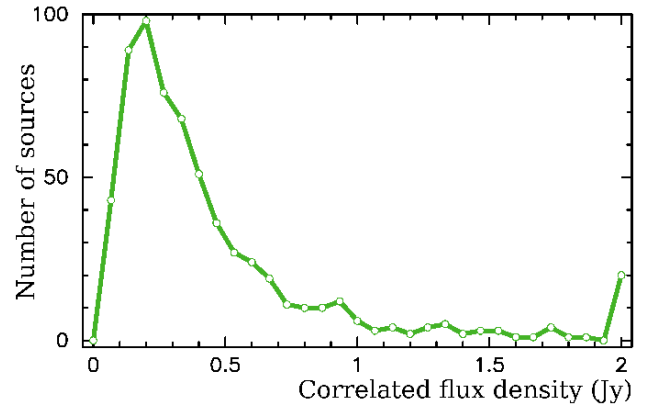}
  \else
     \includegraphics[width=0.46\textwidth]{kcal_distr.eps}
  \fi
\end{figure}

  Figure~\ref{f:distr} shows the distribution of the KCAL correlated 
flux densities. Assuming that the parent population of sources is uniform
in the range of flux densities 1--1000~mJy, we explain the sudden drop 
in the number of sources with correlated flux densities below 200~mJy 
as an indication of under-representation of sources weaker than that limit
in the catalogue because they are not reliably detected with VERA. Thus,
the KCAL is incomplete at flux densities below 200~mJy. This result 
is in agreement with our analysis of previous VERA observations 
\citep{r:vera_fss} where we estimated the probability of detection of 
a source with the correlated flux density 200~mJy at a level of 70\%.

\begin{figure}[h]
  \caption{\ifpre \rm \fi
           The distribution of the spectral indices $\alpha$
           ($F(\nu) \sim \,v^{\alpha}$) of 1536 sources from the input
           catalogue. The spectral index was computed from 
           median correlated flux densities at 8.6 and 2.3~GHz at baseline 
           projected lengths shorter than 900~km.
  }
  \label{f:kspi}
  \par\medskip\par
  \ifdraft
     \includegraphics[width=0.47\textwidth]{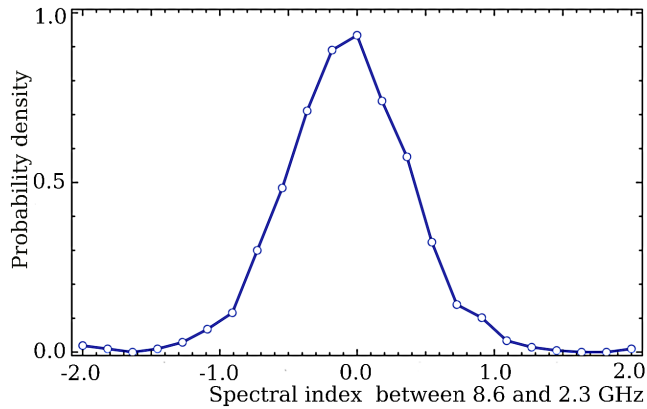}
  \else
     \includegraphics[width=0.47\textwidth]{kcal_spind_distr.eps}
  \fi
\end{figure}

  The detection limit of VERA at 22~GHz, 200~mJy, corresponded 
the lowest correlated flux density of the input source list, 200~mJy,
{\it at 8~GHz}. The majority of the sources from the input list
were previously observed at VLBI at both 8.6 and 2.3~GHz. The distribution 
of spectral indices ($F(\nu) \sim \,v^{\alpha}$) of the compact component
of these sources shows a peak near spectral index 0 (see Figure~\ref{f:kspi}).
Among 1536 sources from the input list, 48\% had spectral index greater
then zero, and therefore, their extrapolated flux density at 22~GHz was 
greater than 200~mJy, the average detection limit of the KCAL survey. 
Although  a measured correlated flux density at 22~GHz for an individual 
source may be less of greater than the flux density extrapolated 
from 8.6/2.3~GHz, if to consider the entire population as a whole, 
the measured flux density at 22~GHz turned out {\it on average} very close 
to the extrapolated one.

  Results of the KCAL survey augmented with results of prior K-band 
surveys form the list of objects with known correlated flux densities. 
By June 2011, this list\footnote{Available at \web{http://astrogeo.org/rfc}} 
contained 1161 objects. Among these sources, 766 objects have correlated 
flux densities greater than 200~mJy at baselines shorter than 70~M$\lambda$
and 608 objects are brighter than 200~mJy at baselines longer 
than 100~M$\lambda$. These sources are considered as a pool of calibrators
for VERA in 2011. After completion of a planned sensitivity upgrade,
even weaker sources can be used for calibrators.

  We reserve a rigorous population analysis to a future publication.
Preliminary results indicate that the number of compact extragalactic
sources at K-band brighter than a given correlated flux density level
is twice less than at the X-band.


  We would like to thank Alan Fey for making publicly available not only
contour plots of images from the KQ survey, but brightness distribution
and calibrated flux densities in the FITS-format. The availability of this
information was crucial for our project.

\end{document}